\documentclass[original]{adada_enhanced}

\usepackage{graphicx}
\usepackage{amsmath}
\usepackage{color}
\usepackage{hyperref}
\usepackage{xurl}
\usepackage[T1]{fontenc}
\usepackage{tikz}

\begin{document}

\title{Access-, Use-, Sustain-ability: Exploration and Quantification of a Linguistic Trend and Its Impact on Design}
\title{\{Access|Use|Sustain|\_\}-ability: Exploration and Quantification of a Linguistic Structure and Its Use in HCI}



\author{Uhde, Alarith}
   {Ritsumeikan University}
   {uhde@fc.ritsumei.ac.jp}

\author{Haimes, Paul}
   {Ritsumeikan University}
   {haimes@fc.ritsumei.ac.jp}

\begin{abstract} 

Language plays a crucial role in research and technology design, throughout problem
formulation, method descriptions, analysis of findings, and interpretation. In
this paper, we focus on a particular, trending linguistic pattern in the
Human-Computer Interaction literature, which we call
\emph{-abilities}. Examples include usability, accessibility, and
explainability. These -abilities have become increasingly popular, with
potential impacts on the discipline. They take a verb, turn it into an
adjective, and
finally nominalize it.  We approach the -abilities phenomenon with a survey of
the HCI literature between 1988 and 2025. All in all, we identified
349 different -abilities. Moreover, the use of -abilities notably increased
since the late 2000s, both in the absolute number of different -abilities and
the frequency of use of the most popular ones. The majority of -abilities we
found had a positive notion, and most -abilities with a negative notion are
prefixed counterparts of their positive opposite (e.g.,
\verb|un+sustainability|). We critically discuss implications of their increased
use. First, -abilities change the literal meaning of design intentions. For
example, \emph{acceptability} reduces the concrete verb \emph{accepting} into
only its potential (\emph{acceptable}), and finally nominalizes it, claiming the
existence of acceptability as a somewhat stable and real, abstract
characteristic of a technology. Second, the prefix-based opposite construction can weaken the design goal. \verb|Un+sustainability| avoids clarification of the
-ability's meaning (here e.g., ecological or financial), and does not spell out
the expected impact of designs that are not sustainable (e.g., the destruction
of nature). Future work should compare the use of -abilities in English with
other languages to better understand the ways in which they transform meaning.

\end{abstract}


\maketitle

\section{Introduction}

The words we use to describe design goals can be understood as linguistic
interfaces between design intentions and product implementations. We might for
example intend to design a website with good \emph{usability}, and then
implement certain features so that visitors experience no major issues.
Likewise, a tote bag intended to be \emph{sustainable} will have features to
protect the environment, and \emph{socially acceptable} wearables will be built
to make people feel comfortable about them in public. In other words, design
goals such as usability, sustainability, and (social) acceptability constitute
an important first mental step that shapes the later design direction.

We are in principle free to pick or create any words to define our design goals,
with perhaps linguistic abilities, creativity, and grammatical rules as the only
limits. In practice, community conventions can serve as additional guidance and
establish commonly used terms, as in the case of \emph{usability}. These help
communication, as they (literally) establish a shared language, which in
turn helps designers identify commonalities in their work. Over time
and across many projects where people use different words to describe their
goals, we can look for patterns of language use, which can tell us something
about our field. For example, by analyzing the terms we can learn about general
interests, trending topics, or shared values, but also potential blind spots or
other issues expressed through the collective use of language.

\begin{figure}[b]
\centering
\includegraphics[width=\linewidth]{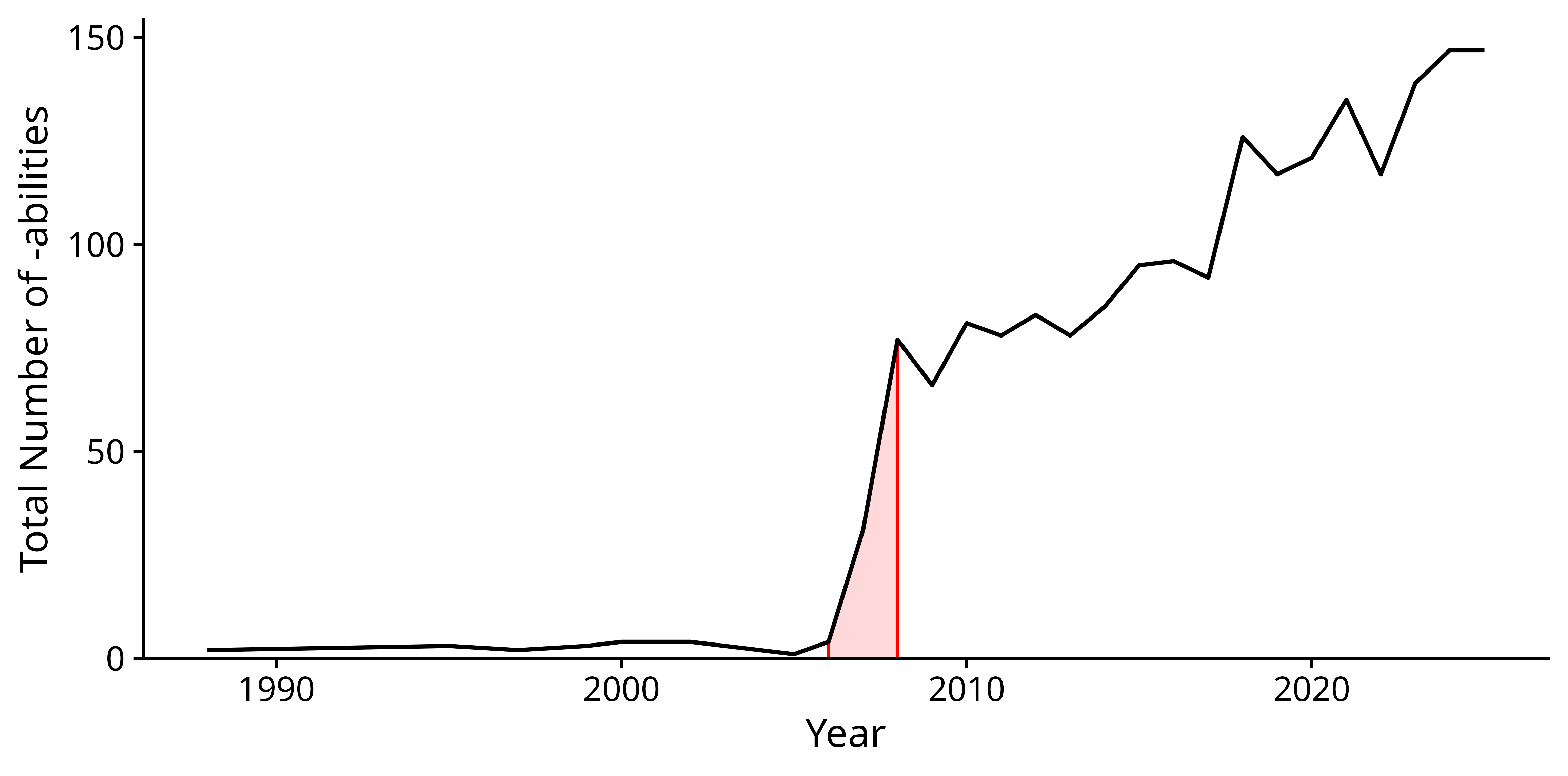}%
\caption{The total number of different -abilities found in HCI paper abstracts per year. The red area marks the years 2006 to 2008 with the sharp increase.}
\label{fig:peak}
\end{figure}

The motivation for this paper grew out of the observation of such a collective
use of language in the Human-Computer Interaction (HCI)
literature. Rather than a specific term, we noticed a grammatical pattern that
has become a trend in the late 2000s and has further risen in popularity ever
since (see Figure~\ref{fig:peak}). We call them ``-abilities'', and we have
already introduced some above. Usability~\cite{ISO9241-11}, accessibility
(e.g.,~\cite{Maia2018}), and sustainability (e.g.,~\cite{Nurjanah2018}) are
among the most popular. In Artificial Intelligence (AI) related research, we
find explainability~\cite{Chazette2021}, contestability~\cite{Hirsch2017}, and
trustability~\cite{Jannel2026}. And in the sciences more broadly, we speak of
generalizability, reliability, and replicability
(e.g.,~\cite{OpenScienceCollaboration2015, Ghai2025}). We have collected several
hundreds of such -abilities from HCI-related papers published over
the past decades. Here, we present our analysis of their use and critically
discuss it.

In sum, the paper makes the following three contributions. First, we confirm and
quantify the widespread use of -abilities in the HCI literature, with
detailed descriptions of their frequencies and developments over time. Second,
we discuss potential explanations for their increasing popularity. And third, we
reflect on the impact of -abilities on the meaning of design goals, along with
the risks and opportunities that go along with that.

\section{Background}

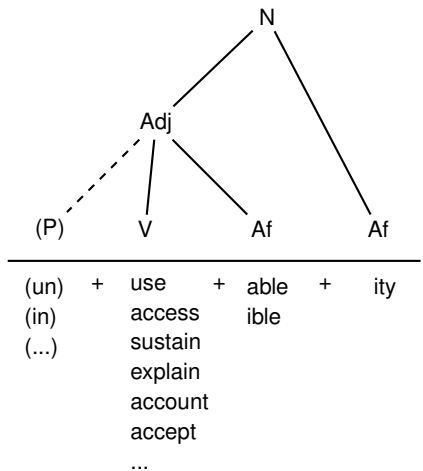
\begin{figure}[b]
    \centering
    \begin{tikzpicture}[
        scale=0.7, 
        transform shape,
        every node/.style={font=\sffamily\Large},
        thick
    ]   

    \node (N) at (2.1, 0) {N};
    \node (Adj_int) at (0, -2) {Adj};
    \node (Af) at (4.2, -4) {Af};
    \node (P) at (-2, -4) {(P)};
    \node (V) at (-0.2, -4) {V};
    \node (Af_leaf) at (2, -4) {Af};
    
    \draw (N) -- (Adj_int);
    \draw (N) -- (Af);
    \draw[dashed] (Adj_int) -- (P);
    \draw (Adj_int) -- (V);
    \draw (Adj_int) -- (Af_leaf);
    
    \draw (-2.8, -4.6) -- (5.0, -4.6);
    \node[align=left, anchor=north west] (Col1) at (-2.6, -4.8) {(un)\\(in)\\(...)};
    \node[anchor=north] (Plus1) at (-1.1, -4.8) {+};
    \node[align=left, anchor=north west] (Col2) at (-0.6, -4.8) {use\\access\\sustain\\explain\\account\\accept\\...};
    \node[anchor=north] (Plus2) at (1.2, -4.8) {+};
    \node[align=left, anchor=north west] (Col3) at (1.6, -4.8) {able\\ible};
    \node[anchor=north] (Plus3) at (3.2, -4.8) {+};
    \node[align=left, anchor=north west] (Col4) at (4.0, -4.8) {ity};
    
    \end{tikzpicture}
    \caption{The root verb (V) of an -ability is expanded with the \emph{able}
      affix to form an adjective (Adj), which is then nominalized (N) with the
      additional \emph{ity} affix (Af). Some -abilities are further modified with a
      prefix (P).}
    \label{fig:wordtree}
\end{figure}

We can characterize -abilities based on their linguistic structure to have a
clearer definition for further analysis. Generally, \mbox{-abilities} are based on a
root verb, which provides them with their core lexical meaning. This root verb
is expanded with the \verb|-able| or \verb|-ible| affix into an adjective, and
finally nominalized using the \verb|-ity| affix. For example, the verb
\verb|sustain| is extended into the adjective \verb|sustainable|, and finally
nominalized into \verb|sustainability| (see also~\cite{Bender2013}). Some -abilities also have a prefix
(e.g., \verb|un+sustainability|).  Depending on the context, the adjective form is also used, although our focus here lies primarily on the nominalized
variant. Figure~\ref{fig:wordtree} illustrates this basic structure.

The -ability suffix also appears in some other words, which we treat separately
here. For example, \emph{ability} is a special case that has no separate root
verb. Moreover, a few other words like \emph{compatibility},
\emph{equitability}, and \emph{palatability} have no root verb either, although
they are grammatically valid. We exclude these words without root verb from our
definition. Finally, some -abilities are based on a verb but use it in an
altered form. Examples include credibility, which comes from the Latin verb
\emph{credere}, (``to believe''), and vulnerability (from Latin
\emph{vulnerare}, ``to wound''). These are included in our definition.

\subsection{The Role of Language in Research and Design}

We found no prior work with a particular focus on -abilities, but critical
reflection on the use of language in general, and its role in shaping our
thinking, has been a key concern at least since the linguistic turn in
philosophy~\cite{Rorty1970}. It was also foundational for postmodern theory
(e.g.,~\cite{Frank1984, Wittgenstein1953}). Over time, this increased attention
to language use has found its way into HCI and design research, notably through
the ``semantic turn''~\cite{Krippendorff2006} and in feminist and postcolonial
perspectives~\cite{Ozkazanc-Pan2012}.

A prominent example is the term ``Artificial Intelligence'' or ``AI''. Over the
decades, this term has undergone critical analysis because of its imprecise and
shifting meaning. Although initially used to describe problem-solving and expert
systems, it has shifted to describe statistical and data-driven models today,
sometimes also chatbots more specifically~\cite{Suchman2023}. ``AI'' is also
commonly used to rebrand existing technologies (e.g., ``Big Data''), or to hide
the human labor the systems are based on~\cite{Suchman2023, Nemer2026,
  Tubaro2020}. This function as a ``floating signifier'' with a drifting meaning
and ``strategical vagueness''~\cite{Suchman2023} serves today as a marketing
tool in politics, business, and research. Besides ``Artificial Intelligence'',
similar critiques have also been voiced about intelligence as such, with its
comparable vagueness and practical consequences, along with attempts to redefine
it~\cite{Attard-Frost2023}.

Unlike ``AI'', -abilities are not a particular term but a broader morpheme
structure that has become popular over the past two decades. While the use of
-abilities in HCI specifically has not been studied yet, such trends
in morpheme structures are not new.

We can find practical examples, such as the trending company names in the late
2000s that omitted a vowel as a means of marketing and recognition
(e.g., flickr, tumblr, or grindr)~\cite{Pathak2022}. Another example was Apple's
former product naming scheme which started with an \verb|i| followed by a
capital letter (e.g., iTunes, iPhone, iPod). This pattern has later been
mimicked by other companies (e.g., iRobot, iHealth, or iZotope), which may have
contributed to Apple's shift away from it. A final example is the ``Smart X''
trend (e.g., smartphone, smartwatch, smart city), which has even sparked its own
``dumb'' countermovement~\cite{Rothschild2024}.

Some research has focused on suffix use in scientific language. For example,
Montero-Fleta's work~\cite{Montero-Fleta2011} compared nominalizations in
scientific language corpora, specifically from medicine and computer science,
with more broadly used English language from the British National Corpus
(BNC). The analysis revealed a high prevalence of nominalizing suffixes
including \verb|-ion|, \verb|-ness|, and \verb|-ist| in the scientific corpora,
compared with the BNC. A second, more recent work by Zagà~\cite{Zaga2025}
studied the use of the \verb|-logy| suffix in its use to name scientific
disciplines. The \verb|-logy| suffix implies scientific authority but is at
times misused to add a scientific appearance to non-scientific or
pseudoscientific contexts (e.g., astrology or phrenology).

\begin{figure*}[t]
\centering
\includegraphics[width=1.0\textwidth]{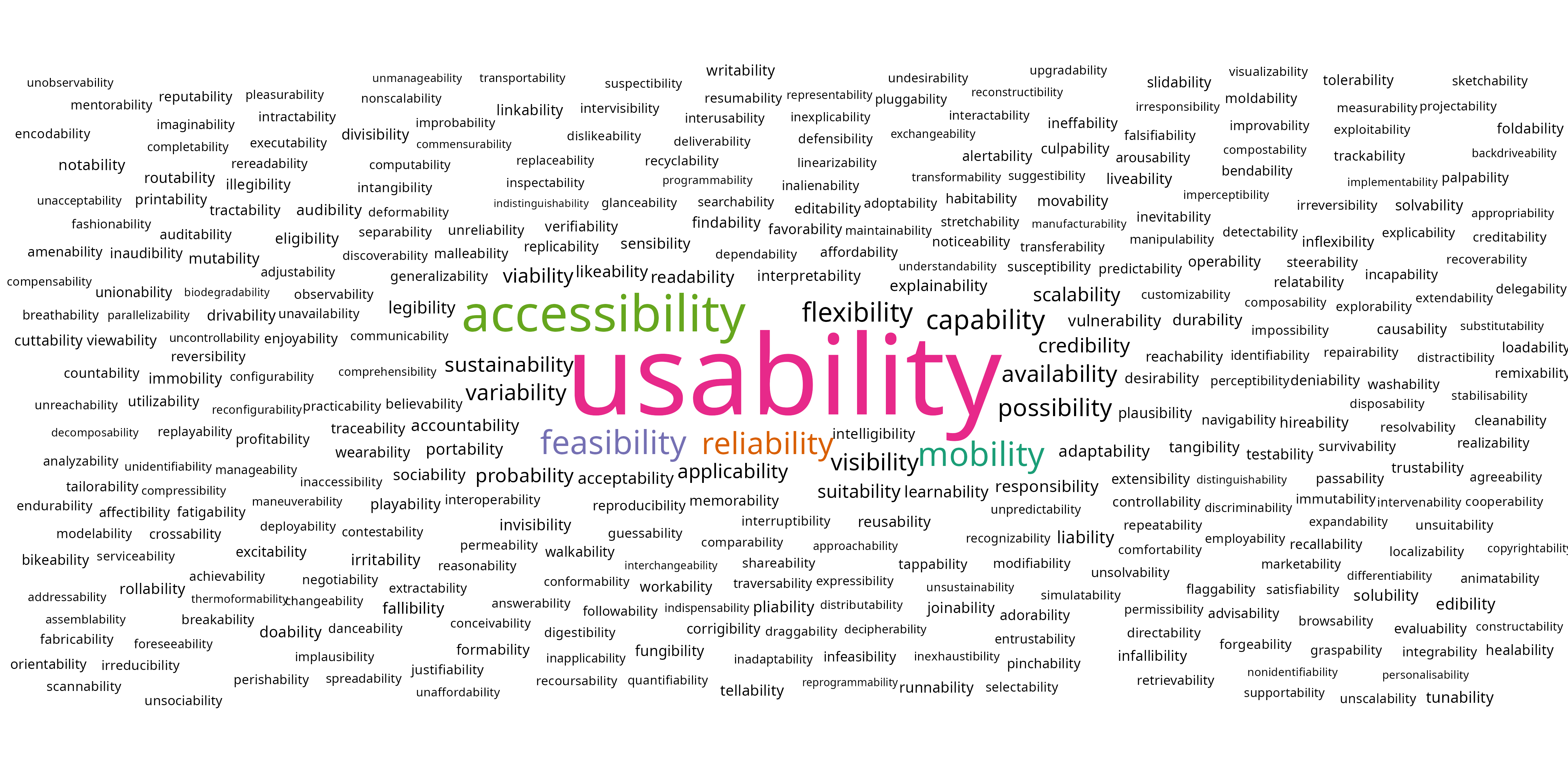}%
\caption{The 349 -abilities used in HCI paper abstracts in the ACM Digital Library, between 1988 and 2025. Text size represents the frequency.}
\label{fig:wordcloud}
\end{figure*}

In sum, critical discussions about language use enrich the discourse and reveal
issues, as in the case of ``AI'' or \verb|-logy|. It also helps describe
linguistic trends both within and outside of academia. One takeaway is that the
choice of words and their form matters. The specific choice of words influences
how we think about design, how we formulate problems or possibilities, and
finally how the further design process develops. To quantify and better
characterize the -ability trend specifically, we conducted a systematic
literature search as outlined below.

\section{Method}

We started by collecting a sample of -abilities from scientific publications in
HCI. In the further process, we quantified their use to identify
which -abilities are most common, how frequently they are used, and how many we
can distinguish. We also identified ``misuses'' in which the -ability
construction was grammatically incorrect. Finally, we looked into the valence
(generally positive or negative meaning of the -ability) and prefix use to
further characterize them.

\subsection{Literature Search}

We ran a search query in the ACM Digital Library (ACM Full Text Collection) for
articles that contain any word ending in \mbox{-bility} in the abstract and were
related to HCI\footnote{The specific steps and scripts can be
found in the supplementary material available at
\url{https://osf.io/2u4sw/overview?view_only=1c4f4e5ee7034e788aeaad6f8069e4e4}}. We
included only research articles, which were published in journals and conference
proceedings before the end of 2025. The ACM Digital Library does not allow the
search for suffixes directly, because it cannot handle search terms that begin
with a wildcard character. Thus, we bracketed the wildcard \emph{*} manually
between each letter of the English alphabet and the \verb|-bility| suffix, which
allows for both the \verb|-ability| and \verb|-ibility| affixes. This led to the
following query:

\begin{quote}
\raggedright\ttfamily\footnotesize
 Abstract:(a*bility OR b*bility OR c*bility OR d*bility OR e*bility OR f*bility OR g*bility OR h*bility OR i*bility OR j*bility OR k*bility OR l*bility OR m*bility OR n*bility OR o*bility OR p*bility OR q*bility OR r*bility OR s*bility OR t*bility OR u*bility OR v*bility OR w*bility OR x*bility OR y*bility OR z*bility) AND AllField:(HCI OR "Human-Computer Interaction")
\end{quote}

After obtaining the list of results, we applied filters based on the publication
date (before 2026) and type (journal or conference proceedings) in the ACM
Digital Library web interface. All in all, this resulted in 23122 hits (3043
journal papers and 20079 conference proceedings publications).

Next, we extracted all -abilities from the abstracts using an automated script,
which resulted in 34523 hits of 420 different words. We then looked through the
420 words and identified several issues, which we verified in the original
abstracts before fixing them. Some of the words included typos (e.g.,
1usability) and some -abilities appeared with different spellings for the same
concept (e.g., customisability and customizability). Moreover, some words were
hyphenated (e.g., [indistin]guishability). We cleaned these by correcting the
typos, standardizing spelling, and completing hyphenated words. After cleaning,
386 distinct words remained.

\begin{figure}[t]
\centering
\includegraphics[width=0.9\linewidth]{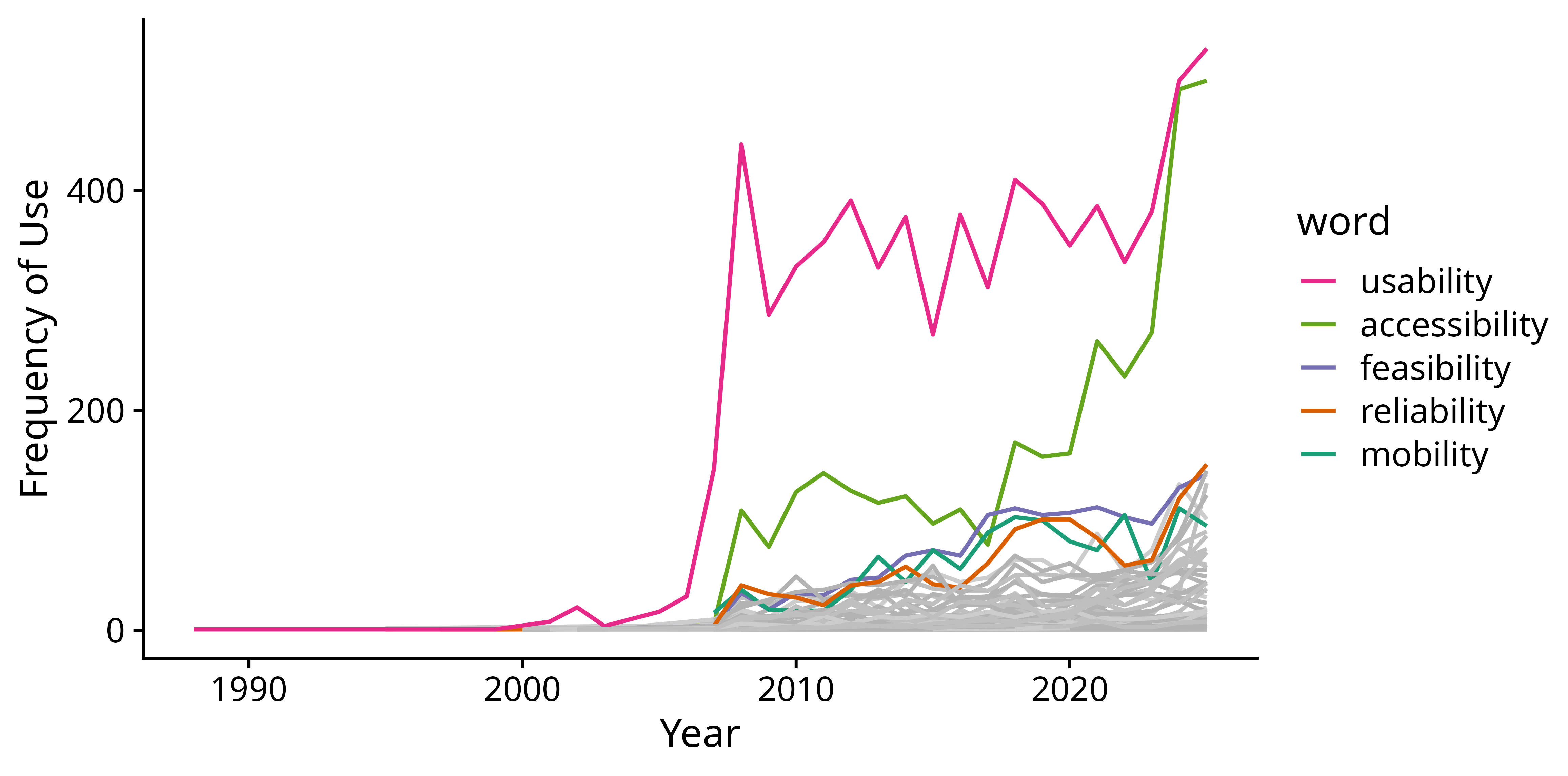}
\caption{Use of different -abilities over time, highlighting the most frequent cases.}%
\label{fig:top_words}
\end{figure}

In the next step, we filtered out words that fell outside the \mbox{-abilities}
definition as described above. Here, we distinguished between two
categories. First, we excluded 12 words that do not contain a root verb but are
grammatically valid (e.g., \emph{equitability}). Second, we excluded 25 further
words that we considered grammatically incorrect, but which seemed to be
intentionally constructed as they are (i.e., they are not typos). One difficulty
of this step is that the existence of a root verb is not always clear. The
English language sometimes allows for nouns to be used as verbs. For example,
one of these cases was \emph{robotability}~\cite{Franchi2025}. Although typically a noun, \emph{to
robot} could also be used as a verb. In such cases, we verified how they were
used in the context of their original abstracts and excluded them if not used as
a verb there. A counterexample was \emph{unionability}, which initially seemed
to use a root noun (union) but was included because it described a table
operation as verified in the original context (\emph{to
union})~\cite{Marimuthu2025}. Finally, although we excluded these grammatically
incorrect cases, we want to stress that we do not see them as an issue in
principle. They indicate the popularity of -abilities, even in unconventional
forms. In a sense, this expresses the impact of -abilities and the range of their creative
appropriation.

\subsection{Exploratory Content Analysis}

In the final step of our analysis, we explored characteristics of the \mbox{-abilities}
themselves with a brief content analysis. Specifically, we quantified a) the use
of prefixes and b) the valence of the \mbox{-ability}.

For prefixes, we considered an -ability as prefixed only if removing the prefix
would leave a valid root verb. For example, the prefix in
\verb|un+sustain+ability| leaves the verb \emph{sustain}, so it would count as
prefixed. Conversely, the prefix in \verb|ad+just+ability|, leaves only
\emph{just}, which on its own is not a valid verb, so this -ability was coded as
not prefixed.

To quantify the valence, we coded each -ability depending on whether achieving
this -ability would in the general case be considered positive, whether the
valence was unclear, or whether it would be considered negative. The goal of
this was to get a broad estimate of how many positive and negative -abilities
are in use, so we based this assessment on an everyday understanding. For
example, \verb|accessibility| would generally be seen as a positive
characteristic of a technology, whereas \verb|irritability| would be
negative. The meaning of \verb|discriminability| is somewhat unclear, because it
can be used both to describe (positive) technical features and (negative) social
impacts. Two authors
coded the -abilities in a first step, with good interrater agreement
($\text{Krippendorff's }\alpha=.83$). In the second step, we discussed and resolved the
remaining 34 -abilities where the ratings differed, which resulted in a final
set of the -abilities' valence.

\section{Results}

\subsection{Overall Number and Frequencies of Use}

After filtering and fixing typos, the final set consisted of 349 valid
-abilities, plotted in Figure~\ref{fig:wordcloud}. Out of these, the most
commonly used -abilities included usability (6984 times), accessibility (3364
times), and feasibility (1442 times). However, the distribution was skewed with
a long tail, and 120 -abilities were only used once. The large difference
between the mean ($82.2$) and median ($3$) reflects this strong skew.

\subsection{Development Over Time}

Figure~\ref{fig:peak} already indicated a strong increase in the number of
\mbox{-abilities}, from 4 in 2006, to 31 in 2007, to 77 in 2008, and an ongoing high
level over the following years up to 147 -abilities in 2025.

In the time before that, between 1988 and 2005, the most frequently used
-ability was usability (52 times), followed by availability (3 times) and a
total of 14 further -abilities only used once or twice. Notably, we found no
-abilities in the ACM Digital Library database before 1988, and in further 9 out
of the 17 years between 1988 and 2006 no -ability was mentioned in any abstract.
The major trends for
the time after the sharp increase that happened between 2006 and 2008 can be
seen in Figure~\ref{fig:top_words}. Usability has remained the most frequent
-ability so far. However, accessibility has also seen a sharp increase since the
mid-2010s and recently reached a high level of 500 uses in 2025, close to
the 529 uses of usability. The following three most frequent -abilities
(feasibility, reliability, and mobility) already have a notably lower usage (in
2025, 142, 151, and 95, respectively).

\subsection{Prefix Use and Valence}

We found 88 -abilities with a prefix, and a clear majority of \mbox{-abilities} with
positive valence (291), followed by negative valence (36), and unclear valence
(22). We also explored the intersection of prefix use and valence, and found
that most -abilities with negative valence (29) had a prefix that negated or
inverted their original meaning (e.g., un-, dis-, in-, or non-). There were only
seven negative -abilities without prefix: fallibility, fatigability,
irritability, perishability, susceptibility, suspectibility, and vulnerability.

The 22 -abilities with unclear valence could be subdivided into three different
subtypes. Four -abilities were relatively objective descriptions and used neutral or technical root verbs (e.g., causability, improbability). Eight used
potentially ambiguous terms with different meanings, depending on the context
(e.g., discriminability, exploitability, manipulability). And ten were prefixed,
inverted -abilities that described the absence of something (e.g.,
intangibility, invisibility, irreducibility).

\section{Discussion}

This paper presented a study on the use of -abilities in the HCI and design
literature. We identified 349 valid -abilities, with a sharp increase of use in
the late 2000s and a steady increase after that. The most frequently used
-abilities were usability, accessibility, and feasibility. Usability has a
historically high usage, whereas accessibility has recently reached similarly
high usage levels. Many other -abilities only have a low usage, with 120
-abilities only used once. Finally, the majority of -abilities has a generally
positive meaning and is not prefixed. Among the \mbox{-abilities} with negative
meaning, the majority is prefixed. A few further -abilities have neutral or
ambiguous meanings.

\subsection{Why Are -abilities Increasingly Popular?}

The -abilities phenomenon is still relatively recent, with very low numbers
before 2006 and a sharp increase after that. The particular timing coincides
with a few major shifts in the technology landscape we outline here. However,
our analysis does not allow for causal attributions, and we assume that a
combination of multiple aspects may have led to the phenomenon. Moreover, the
sharp increase in the late 2000s both happened in the use of usability as the
top -ability (see Figure~\ref{fig:top_words}) and in the total number of
-abilities (see Figure~\ref{fig:peak}).

The first major shift was described by Susanne Bødker in her article on the
third wave of HCI~\cite{Bodker2006}. In the 1980s and 1990s, research had mostly
focused on office contexts with single users (first wave) and groups (second
wave). Accordingly, the design goals were mainly focused on efficiency, user
errors, and group coordination. In the third wave, the interest in other domains
of computer use increased, and a broader set of design goals went along with
that (e.g., joyful user experiences~\cite{Hassenzahl2006a}). Although this
increased breadth of applications and design goals could have contributed to
the number of -abilities in use, Bødker's paper described a process that was
already happening before 2006, so slightly before the effect shown in our data.

The second and related shift was the increasing variety of mobile devices. The
first iPhone was released in 2007, but other mobile technologies such as mp3
players were already widely in use. In the following years, more devices such as
tablets, fitness trackers, and smartwatches became widely available. Mirroring
the third wave trend, these products broadened the variety of contexts and
applications of interactive technology. This variety may have gone along with an
increase of -abilities to describe new design goals.

We have considered further explanations, but some of them seem less likely to have
contributed to the sharp increase. For example, it would be possible that a new, popular
measurement tool for usability was published briefly before. However, the most
widely used measure for usability today, the System Usability Scale (SUS), was
already published in 1996~\cite{Brooke1996} and the ISO standardized definition of
usability in 1998~\cite{ISO9241-11}. A further issue could be changes made to the
ACM Digital Library database or a general increase in the number of publications
over the years, both of which we could not verify or rule out with our data. In
addition, we looked for trends within our dataset, to test whether changes over
time in the frequency of -ability use per abstract, the variety of -abilities per 
abstract, or abstract length may have contributed to the effects. We found negligible
correlations for the total number of -abilities (Spearman's $\rho=-.007$) and the
number of different -abilities per abstract (Spearman's $\rho=.02$, $p<.01$) over
time, which seemed not to have played a major role. However, there was a small
increase in abstract length over time (Spearman's $\rho=.22$, $p<.001$), which may
have contributed to but not fully caused the effect. Further details can be found in the
supplementary material. Finally, there is the possibility of self-reinforcing effects:
Readers may notice the increasing use of -abilities in the literature, which
could suggest it as a candidate for new word creations.

\subsection{What Work Do -abilities Do?}

So far, we have discussed external reasons that may have led to the increased
use of -abilities over the past years. Here, we want to focus on internal
reasons or ways in which the particular morpheme structure of -abilities shapes
the meaning of the underlying concepts, and how that might be particularly
attractive for HCI researchers.

We illustrate these effects below with concrete examples that link the
linguistic notions to their impact on thinking about design. However, we
would like to point out that the linguistic, ``literal'' meaning is not the only
meaning of a word, and that it is not always defining for the concept it
describes~\cite{Bender2013}. For example, the word \emph{usability} has a
precise definition~\cite{ISO9241-11}. This definition includes aspects such as
user satisfaction and a context of use, which do not appear in the word
itself. Someone unfamiliar with the usability definition cannot derive these
aspects by only reading the word.

On the other end of the precision spectrum, the word ``sustainability''
encompasses multiple, ambiguous, and partially contradictory meanings.  It can be
used in the context of environmental protection and efforts against climate
change. However, it can also be used in the context of ``sustainable
development'' to argue for economic growth, which can have opposing
effects~\cite{Brown2016a}.

That said, a look at the literal meaning can still be insightful to critically
reflect on the use of -abilities. Following our definition, -abilities take a
root verb, turn it into an adjective, and finally nominalize it. These
transformations on the linguistic level correspond to transformations on the
conceptual level.

\subsubsection{-abilities Imply the Existence of a Real Phenomenon}

Take the example of a wearable device that is designed to have high social
\emph{acceptability}. The first and perhaps strongest claim already made at this
point is that something like ``social acceptability'' exists, as a potential
characteristic of an object, which may be measurable and assumably a somewhat stable
latent construct. An implication of this claim is
that we can take two devices or forms of interaction, one with higher and one
with lower social acceptability, and we would generally find one of them to be
more acceptable than the other.

However, this relation is less clear in practice. As Uhde and
Hassenzahl~\cite{Uhde2021EA} pointed out, the acceptability of a ``phone call''
varies significantly across contexts. The same phone call can have a low
acceptability when performed during a classical concert, and high acceptability
in a shopping mall---even if the performance itself is identical. But the form
of interaction also has inconsistent impacts. Some early work has found that
``subtle'' gestures (i.e., hidden in natural movements) have high acceptability,
whereas ``suspenseful'' gestures (i.e., visible input gestures with
invisible effects)~\cite{Rico2010a, Reeves2005, Montero2010} have low
acceptability. However, later work found not only that suspenseful gestures can
also have high social acceptability~\cite{Uhde2025b}, but that the experience of 
suspenseful gestures can even be better than for subtle gestures, for example in
face-to-face interactions~\cite{Uhde2022d}. In other words, the notion of
``acceptability'' is highly contextual and may not be usefully described as
a characteristic of a technology or form of interaction alone.

\subsubsection{-abilities Transform Action Into Potential Action}

The second step made by -abilities is the addition of the \verb|able|
affix. This turns a verb into an adjective. At the same time, it transforms the
action itself into its potential. For example, \emph{to accept} becomes
\emph{acceptable}. In the verb form, we make a stronger assertion that people
actually accept a technology. In the adjective form, the technology only needs
to be potentially acceptable---not actually \emph{accepted}. This frees the
designer from the onus of proving that they have achieved their design goal. If
we observe that users do not accept a technology, it is (objectively) not
accepted. However, it may still be (potentially) acceptable, even if the
accepting has not happened for (perhaps external) reasons. In other words, this
transformation blurs the conditions under which we claim that the technology
would be accepted, and failure to meet them cannot be accounted to the designer
anymore.

\subsubsection{-abilities Abstract and Recontextualize}

Another shift that goes along with the transformation of action into potential
action is the omission of who or what is acting, and who or what is being acted
upon. In the verb form (\emph{to accept}), we need to make explicit who is
accepting what. Users may accept the technology. Other people may accept that
users use the technology. However, by transforming the verb into an adjective,
the (potential) accepters become abstract. This issue has led to confusions in
the social acceptability literature, which now contains some work focusing on
users as accepters and other work on other people who witness the interaction as
accepters (e.g.,~\cite{Koelle2020a, Alallah2018, Uhde2025a}). But the same issue of
abstract users affects other -abilities as well. In the case of
\emph{sustainability}, it is often not clear who is sustaining exactly
what~\cite{Brown2016a}.

An alternative reading of this abstraction is a shift of focus, away from
characteristics of the actors and towards the interaction itself. Put
differently, the actual ``Human'' and ``Computer'' in Human-Computer Interaction
become secondary concerns, and the ``Interaction'' moves into the center. This
position can in part be supported if we look into the specific -abilities from
our collection. Several examples can be seen as more specific sub-forms of
usability. For example, we found \emph{draggability}, \emph{browsability}, and
\emph{foldability}, each of which defines a more specific action than \emph{to
use}. Thus, the loss of specificity of actors may go along with an increasing
specificity of the interaction. The sheer number of -abilities in use today also
supports that notion.


\subsubsection{-abilities Could Benefit From Clearer Opposites}

One of our findings was that most -abilities have a positive connotation. This
makes sense, given their use as design goals. However, we also found several
-abilities with negative connotations, which were mostly constructed by taking a
positive -ability and negating it with a prefix. For instance, we found
\emph{unsustainability} as an opposite of \emph{sustainability}, and
\emph{inaccessibility} as an opposite of \emph{accessibility}.

Given that our study only focused on -abilities, our corpus is somewhat
incomplete. For example, we cannot see whether other papers have spelled out
what they consider to be the opposite of \emph{sustainability} beyond
\emph{unsustainability} if they used words with other endings. That said,
we did find several prefix constructions in
use to negate -abilities, which prompted us to reflect on this usage.

Specifically, we have two concerns with prefixed negations of \mbox{-abilities}.
First, they do not spell out what they mean. Describing a system as
\emph{unsustainable} tells us little about its effects. This lack of a clear
opposite may contribute to its ambiguity. If we talked about environmentally
unsustainable systems by instead describing them as environmentally destructive
or ecocidal, the environmental framing would be clear. This would also strengthen 
our case to design for sustainability, because we learn about the negative
alternative.

Second, prefixing sometimes alters the root verb into a form that makes no sense
as a standalone verb anymore. The verb \emph{to sustain}, although abstracted
away, can still be interpreted after we clarify the actors. But the (non-)verb
\emph{to unsustain} cannot. The reason for this is that prefixes sometimes
negate the verb itself (e.g., in \verb|dis+cover|), and sometimes the adjective
(e.g., in \verb|un+sustainable|). This further removes the -ability from its
potential action. To avoid such issues, we recommend unwrapping
negated -abilities into the prefixed verb form and using alternatives if the
verb makes no sense on its own.

\subsection{Future Work and Limitations}

This paper has provided a first critical reflection on -abilities, and we
hope that this sparks a debate on their use and potential alternatives in the
future. However, it also comes with several limitations.

First, the corpus we analyzed is limited in several ways. It only included articles
with abstracts that have words ending in -ability. This is the nominalized form, and
nominalization may be used differently in abstracts compared with full texts of articles (e.g.,
to reduce word counts). Thus, the abstract-based analysis may not fully generalize to
the full articles. Likewise, because of the -ability filter, we could not conduct a comprehensive analysis of the opposite pairs, to see if they may have been constructed 
with different suffixes. The article database itself adds further restrictions. We only 
used publications covered in the ACM Digital Library. This excluded various venues from
adjacent fields, such as design research. And finally, the keyword-based approach to
filter for HCI-related articles may have been imprecise in some cases, and may have
added a bias to the dataset that is different from for example venue-based filters.

In addition, we restricted the corpus analysis on descriptive quantifications.
More in-depth, qualitative analyses of particular -abilities in their use contexts may 
reveal further issues or potentials. Finally, the discussion focused on the literal 
meaning of the -abilities~\cite{Bender2013}, independent of alternative definitions
that exist in some cases.

For future work, we would be interested in comparative analyses with translations
used in other languages. For example, the double function of nouns as verbs in English 
might cause language-specific issues that take different forms in other languages.

\section{Conclusion}

The linguistic phenomenon of -abilities shapes how we conceptualize design
intentions, and how we think about the steps to implement them. -abilities have
become increasingly popular since the late 2000s, both in the total number of
-abilities and in their frequency of use (e.g., usability or accessibility). We
identified several opportunities for reflection, including the claim of real
existence of a phenomenon, the transformation of actions into potential actions,
their abstraction away from concrete actors and contexts, and the lack of
spelled-out opposites.

We invite HCI researchers and designers to reflect on their use of -abilities when
formulating their design intentions. When designing a new tote bag, for example,
rather than selecting ``sustainability'' as an abstract intention, designers could
try to specify who should do the sustaining of what exactly and in what way.
Likewise, the \emph{social acceptability} of wearables should be specified more
precisely in terms of who, in which contexts, does the accepting. In that sense, we
trust in the reflectibility, reconsiderability, and finally the flexible adaptability
of our field's language use.

\section*{AI Acknowledgments}

LLMs were used to convert the format of Figure~\ref{fig:wordtree} from png into
tikz to increase accessibility, and for coding assistance for the scripts
provided in the supplementary material, with human oversight. No LLMs were used
for writing or conceptual work.

\bibliographystyle{adada-refs} 
\bibliography{references.bib}

\end{document}